# New Worlds:
# Evaluating terrestrial planets as astrophysical objects

Submitted as a white paper to the 2010 Astronomy & Astrophysics Decadal Survey

Primary Author: Caleb A. Scharf,
Columbia Astrobiology Center, Columbia Astrophysics Laboratory, 550 West 120$^{th}$ St., MC 5247, New York, NY 10025. Tel: 1-212-854-4451, email: caleb@astro.columbia.edu

Co-Authors: David S. Spiegel (Department of Astrophysical Sciences, Princeton), Mark Chandler (Center for Climate Systems Research,Columbia), Linda Sohl (Center for Climate Systems Research, Columbia), Anthony Del Genio (NASA/GISS), Michael Way (NASA Ames/GISS), Nancy Kiang (NASA Ames/GISS)

Thematic Science Area: *Planetary Systems and Star Formation*

# Introduction

One of the particular motivations for exoplanetary astronomy is the search for terrestrial worlds, potentially capable of harboring life. Within the next five to ten years hundreds, perhaps thousands, of these small rocky planets will probably be discovered. This represents an extraordinary scientific opportunity for astronomy, planetary science, and even biology. It also presents an enormous challenge. In a few special cases, spectral data that probe the atmosphere of terrestrial worlds might be available - specifically those transiting low-mass stars, using for example *JWST* (Charbonneau & Deming 2007). Interpreting such data will be extremely difficult. The signatures of key molecules such as $H_2O$ or $O_2$ will need to be taken in the context of the overall environmental condition of a planet – which may be very different from that of the Earth. Although observational constraints on these worlds might be quite limited, the astronomical community will still need to evaluate their potential characteristics and capacity to harbor life, commonly described as their habitability. This will be necessary both to place the Earth in proper context (is it rare or common?), and to critically inform future decisions on astronomical instruments and missions designed to seek detailed measurement of these planets (e.g. Terrestrial Planet Finder concepts).

This white paper presents a strategy for answering the central observational questions about terrestrial planets that face astronomy:

1. What are the surface and atmospheric conditions ?
2. Does the planet represent a plausible environment for life now, in the past, or in the future ?
3. What is the overall compositional nature of the planet ?
4. Is there evidence for the presence of an active geosphere ?
5. Is there evidence for the presence of an active biosphere ?

Foremost are issues of the surface and atmospheric environment of these worlds. The expected diversity of formation histories, orbital characteristics, stellar parents, and geophysical evolution open up an enormous parameter space. Astronomy cannot work in isolation on the observation and interpretation of terrestrial planets. Already, there is increasingly interdisciplinary activity with planetary science, geophysics, and biology that attempts to bridge the gaps. The formal support for such efforts is enormously important. Even if detailed measurements are available for a planet, modeling something as fundamental as climate[1] is presently extremely difficult. In part, this is due to the inherently complex nature of climate, but it is also due to the understandably Earth-centric development of climate theory and models. State-of-the-art Earth climate models cannot easily be re-tuned for arbitrary planets; even modest changes to the spin rate of a planet can effectively crash such simulations. The strategy discussed here requires the development of a fundamental modeling hierarchy to understand terrestrial planet climates, and observables, as needed by astronomers. This hierarchy begins with the simplest physically meaningful models for planetary climates, and leads to more complex, general climate models that can even incorporate the possible effects of a biosphere. With a hierarchy of models, the underlying

---

[1] "Climate" in this context follows the formal definition: the totality of the atmosphere, hydrosphere, geosphere, and biosphere and their interactions

physical relationships between – for example - planetary orbital eccentricity and climate can be studied. This is critical, given the enormous parameter space that can be explored, and that may indeed be represented by real planets. This strategy will help astronomy to move beyond the planetary science of our solar system to the planetary astrophysics of the Galaxy.

## Terrestrial planets as astrophysical objects

The dynamics of energy transport are critical in understanding the environment of a given terrestrial planet, and interpreting its atmospheric chemical equilibria. Because the Earth has always served as the primary case study, the overwhelming majority of climate models (simple and complex) are highly Earth-centric. Changing seemingly trivial parameters (e.g. day length) can be fraught with problems in these models. Even extending simple climate models, such as those based around radiative energy balance and atmospheric photochemistry, to new regimes is non-trivial. Nonetheless, it is likely that terrestrial exoplanets will occupy a diverse range in parameter space – from orbital and spin configurations to varying stellar parents and composition, including key factors such as surface water content.

Proposed here is a logical and practical framework that could be used to develop both a physical understanding of the energy transport regimes of terrestrial exoplanets, and to construct the next generation of climate models – capable of dealing with situations not represented in our solar system. Ultimately, this all leads to a capacity for astronomy to correctly interpret observations of terrestrial planets and to predict the requirements for future observations.

The transport of surface and atmospheric energy on a terrestrial planet is primarily driven by astronomical forcings: stellar insolation, orbital configuration, obliquity, and spin period. Secondary elements include geophysical activity and tidal phenomenae. Only some of these factors will, in general, initially be known for terrestrial exoplanets. A robust, physically meaningful, methodology will be critical in evaluating the nature of any given world. This can be developed through the use of a ***hierarchy of climate models***, from simple 1-D (latudinally resolved) energy balance models that incorporate basic or complex (photochemical) radiative transport, to 2-D energy balance models (suitable for examining energy transport away from persistent sub-stellar points, such as in tidally locked planets), all the way up to increasingly complex full 3-D general circulation models (GCMs) for terrestrial planet atmospheres that include hydrological cycles, realistic radiative transfer (Miller-Ricci, Seager, Sasselov 2009; Showman et al. 2008), and both passive atmospheric chemical tracers and realistic treatment of dynamic chemistry. Given the enormous range of possible external conditions, forcing, and physical processes that can affect planetary climates, it is infeasible to simulate every possible configuration. Computationally-efficient 1D and 2D energy balance models with parameterized dynamical heat transports can be used to broadly survey the planetary parameter space. Flexible general circulation models that allow for explicit simulation of the 3-dimensional dynamics of dry atmospheres, but with simplified parameterizations and a limited number of free parameters, can then be used to study in more detail the atmospheric transport properties of terrestrial planets under a subset of astronomical forcing conditions informed by the lower level models.  Finally, a full terrestrial planet  3-D general circulation model – capable of capturing hydrological cycles - should be developed for astronomy in order to examine specific, and restricted cases. Table 1 summarizes a possible modeling hierarchy.

| Model | Dimensions | Atmospheric composition | Dynamics | Radiative transfer | Chemistry | Land-ocean distribution | $H_2O$ cycle |
|---|---|---|---|---|---|---|---|
| **1-D Energy Balance** | 1 | Implicit | Parameterized | Implicit | Implicit | Implicit in albedo and thermal inertia | No |
| **2-D Energy balance** | 2 | Implicit | Parameterized | Implicit | Implicit | Explicit in albedo and thermal inertia | No |
| **Flexible General Circulation Model** | 3 | Implicit in radiative properties | Explicit | Implicit | Implicit | Explicit in albedo and thermal inertia | No |
| **Full Terrestrial Planet General Circulation Model** | 3 | Explicit | Explicit | Explicit | Explicit | Explicit | Yes |

The calibration of the model hierarchy is a serious and critical issue. Our solar system presents a very limited number of cases to utilize. However, the *paleo*-climate of the Earth (and to a more limited degree Mars and Titan) offers a unique and potentially important test for models of terrestrial worlds similar, but not identical, to the modern Earth. It is therefore proposed that formal support should be given to connecting the knowledge-base of paleo-Earth to any modeling effort for exoplanets.

**Detailed comments on modeling strategies:**

**(a) Energy Balance Climate Models**
A broad survey of the influence of key global properties on planetary climate is essential to lay the groundwork for the interpretation of discoveries by *Kepler*, ground-based (transit & radial velocity) searches, and in anticipation of direct *JWST* measurements for terrestrial planets around nearby M-dwarfs (Charbonneau & Deming 2007). The simplest useful approach for astronomy utilizes 1-dimensional Energy Balance Models (1D EBM). In the 1D EBM framework, a time-dependent diffusion equation is solved to evolve the longitudinally-averaged surface temperature conditions on a terrestrial planet, throughout its seasonal cycle. A heating function represents the variation in incoming flux with latitude (and time), the local albedo is a function of surface temperature, the infrared cooling function captures the atmospheric greenhouse effect, latitudinal heat transport is approximated as a diffusion process and a surface heat capacity captures the effective thermal inertia of the climate system (e.g., Williams & Kasting 1997; Spiegel, Menou & Scharf 2008, 2009). Ever since the seminal work of Budyko (1969) and Sellers (1969) on the

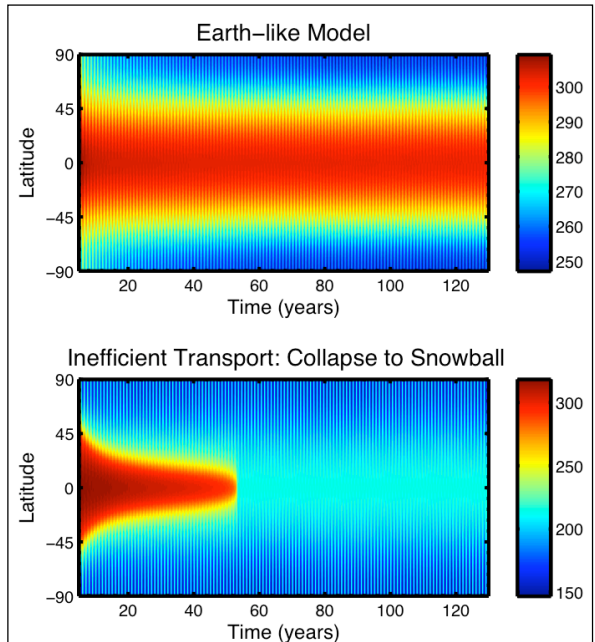

**Figure 1** Latitudinal temperature maps (in K), as a function of model time, in two Energy Balance Models (EBMs) for Earth-like planets at 1 AU from a Sun-like star. *Top*: A fiducial Earth model. *Bottom*: A faster rotating planet collapsing to a snowball, (globally-frozen) state as a result of reduced latitudinal heat transport.

Earth climate, 1D EBMs have been recognized as useful tools for studies in physical climatology, especially with respect to changes in external (astronomical) forcings and their consequences for climate stability (e.g., Hartmann 1994).

Figure 1 illustrates how an EBM study indicates that a planet's rotation rate, via its influence on the efficiency of latitudinal heat transport, can affect its climatic stability against global glaciation ("snowball") events. The upper panel shows the seasonal climatology in a fiducial Earth model (obliquity: 23.5°; ocean fraction: 70% uniformly; orbital distance: 1 AU). To within a few degrees (K), this idealized model successfully reproduces the Earth's seasonal variations of longitudinally-averaged surface temperatures, from pole to equator. Recalling that habitability requires surface temperatures in the range 273-373 K, it is clear from this panel that mid-latitude and polar regions are habitable only for a fraction of the year. The lower panel in Fig.1 shows a model similar in every way to the fiducial Earth-like model, except for an efficiency of latitudinal heat transport reduced by a factor 9, which might correspond to an Earth rotating about 3 times faster (Spiegel et al. 2008). Rather than relaxing to a temperate and regionally habitable climate, this model with reduced poleward heat transport is globally dragged into a frozen snowball state after several tens of years. This type of sudden climate transition has been observed in much more advanced climate models (e.g., Baum & Crowley 2003) and it is in fact believed that Earth itself experienced one or two snowball events in its past (e.g., Hoffman & Schrag 2002).[2]

A great advantage of a 1D EBM approach is that it can be used to systematically explore the climate and habitability of an Earth-like planet as a function of orbital distance from a Sun-like star. The models can be run to determine the orbital region over which a zero-eccentricity Earth-like planet is seasonally (over part of the year) or regionally (over part of its surface) habitable. A 1-D EBM is also remarkably successful at capturing features of high obliquity planets (Spiegel, Menou, Scharf 2009).

---

[2] A terrestrial planet may exit a snowball state by increasing its $CO_2$ atmospheric content, as may have been the case for paleo-Earth (e.g., Hoffman & Schrag 2002). These are slow evolutionary effects. Determining whether or not a planet experiences snowball events can be important for its instantaneous or its past habitability properties as well as its observable characteristics.

## (b) 2-Dimensional Energy Balance Models

The 1D EBM framework implicitly assumes that longitudinally-averaging a planet's climate is a valid approximation as long as the heat fluxes due to transient longitudnally varying eddies are parameterized. While this is justified for the fast-spinning Earth, with its rather slow atmospheric response time, this approximation might fail in many astrophysically interesting situations. 2-dimensional (longitude + latitude) versions of time-dependent EBMs for exoplanets. are necessary if one is to model climate on very slowly rotating planets (for which the assumption of diurnally-averaged insolation breaks down) — in particular planets in the habitable zones of low-mass stars, which would be tidally-synchronized (around M-dwarfs) or slowly rotating (around K-dwarfs). They also allow for explicit modeling of land-ocean configuration impact on heat transport. The development of a 2D EBM requires co-adding several latitudinal 1D EBMs, at various longitude points, and accounting for the larger magnitude of longitudinal atmospheric transport relative to the latitudinal one (which is inhibited by the Coriolis force, especially at high latitude).

## (c) A Flexible 3-D General Circulation Model of Atmospheric Regimes on Terrestrial Planets

A modeling strategy based on simple EBMs is efficient for broadly surveying the vast parameter space of plausible climate regimes and surface habitability conditions. Moving to 3-D models then enables the explicit simulation of the dynamics that are just parameterized in EBMs. More complex climate models can also account for detailed radiative, chemical and hydrodynamical processes. This is particularly true for atmospheric transport, which strongly influences the temporal and regional habitability of terrestrial exoplanets and yet is modeled simply as diffusion in the EBM approach. While Lorenz (1979) provides a justification for the diffusion approximation, developing a more fundamental understanding of how transport varies with global planetary attributes remains a critical aspect of the general study of exoplanetary climates and habitability.

A flexible general circulation model (GCM) should be developed for astronomical studies of terrestrial planets that solves the full primitive equations of meteorology together with a simple treatment of radiative forcing. The adaptation of models from earlier climate work (e.g. Hoskins & Simmons (1975)) would enable this. Such a model would readily permit changes in global planetary parameters (radius, surface gravity, rotation rate, atmospheric gas constants) and would include radiative forcing via Newtonian cooling (a linear relaxation to an assumed equilibrium temperature profile on a specified radiative timescale).

## (d) Terrestrial Planet Global Climate Models – A New Tool for Astronomers to Explore the Environments of Potentially Habitable Worlds

Ultimately, a full 3-dimensional general circulation model, of the kind used to study Earth's climate, is needed. GCMs that incorporate radiative transfer have been adapted to the context of gas giant planets (Showman et al. 2008), where the isentropic atmosphere base simplifies calculations. Terrestrial planet GCMs that include the influence of a solid or oceanic bottom, hydrology, and flexible chemical and radiative transfer prescriptions, will soon be needed, too.

Radiative transfer schemes can be generalized to admit a much wider range of temperatures, pressures and composition, including for example O(1 bar) of $CO_2$, $H_2O$, etc., that might be found on planets within the habitable zone. Versions of GCMs adapted for Mars also include physics for the primary atmospheric constituent to freeze out, leading to local atmospheric collapse (Allison et al. 1999); such physics, which is an issue for tidally locked planets orbiting M-dwarfs (Joshi et al. 1997), is critical. A great deal of work will be needed on feedback processes, in particular hydrological cycles, and the effects of land/ocean configurations on thermal inertia. The computational overhead of these full GCMs will necessarily restrict their use to very specific "slices" of parameter space – in part determined by the simpler modeling hierarchy, and by the requirements of observations of specific objects. If "Rosetta Stone" terrestrial planets are detected and observed, there will be an urgent need for a terrestrial planet GCM flexible enough to be adapted to these cases.

## (e) Early Earth and Early Mars as Analogs for Potentially Habitable Exoplanets

The Earth through time is a wonderful example of how varied the environment of a habitable world can be. Together with Mars, it provides a critical test not only of models for habitable exoplanets but of the identifying characteristics of worlds for which we *know* the potential for Earth-like life can exist, even though planetary surface conditions differ dramatically from the modern Earth. The nature of alternate habitable Earths as revealed by examining certain key geological intervals and explored with fully 3-dimensional general circulation models, is a critical and powerful tool that has not yet been exploited by astronomy. It begins with the earliest Earth and prebiotic conditions and traces the interplay between evolving life and climate from the Eoarchean through Paleozoic eras (~4 Gyr ago to 250 Myr ago).

## Summary

Terrestrial exoplanets are on the verge of joining the ranks of astronomically accessible objects. Interpreting their observable characteristics, and informing decisions on instrument design and use, will hinge on the ability to model these planets successfully across a vast range of configurations and climate forcings. A hierarchical approach that addresses fundamental behaviors as well as more complex, specific, situations is crucial to this endeavor. Incorporating Earth-centric knowledge, and continued cross-disciplinary work will be critical, but ultimately the astrophysical study of terrestrial exoplanets must be encouraged to develop as its own field.